\documentclass[%
 preprint,
 amsmath,amssymb,
 prd,
]{revtex4-1}

\usepackage{graphicx}
\usepackage{caption}
\usepackage{subcaption}
\usepackage{dcolumn}
\usepackage{bm}

\begin{document}

\title{Small-x analysis on the effect of gluon recombinations inside hadrons in light of the GLR-MQ-ZRS equation }

\author{M. Lalung}
\email{mlalung@tezu.ernet.in}
\author{P. Phukan}
\email{pragyanp@tezu.ernet.in}
\author{J. K. Sarma}
\email{jks@tezu.ernet.in}
\affiliation{
 HEP Laboratory, Department of Physics, Tezpur University, Tezpur 784028, Assam, India }

\date{\today}

\begin{abstract}
 We present a study of the contribution of antishadowing effects on the gluon distribution functions $G(x,Q^2)$ in light of the Gribov-Levin-Ryskin-Mueller-Qiu, Zhu-Ruan-Shen (GLR-MQ-ZRS) nonlinear equation at small-$x$, where $x$ is the momentum fraction or Bjorken variable and $Q^2$ is the four momentum transfer squared or photon virtuality. In this work, we have solved the GLR-MQ-ZRS nonlinear equation using Regge like behaviour of gluons in the kinematic range of $10^{-2}\leq x \leq 10^{-6}$ and $5\,GeV^2\, \leq Q^2\leq 100\, GeV^2$ respectively. We have obtained the solution of $G(x,Q^2)$ by considering two particular cases: (a) $\alpha_s$ fixed; and (b) the leading order QCD dependency of $\alpha_{s}$ on $Q^2$. A comparative analysis is also performed where we compare the gluon distribution function due to inclusion of the antishadowing effect with that of the gluon distribution without including the antishadowing effect. Our obtained results of $G(x,Q^2)$ are compared with NNPDF3.0, CT14 and PDF4LHC. We also compare our results with the result obtained from the IMParton C++ package. Using the solutions of $G(x,Q^2)$, we have also predicted  $x$ and $Q^2$ evolution of the logarithmic derivative of proton's $F_2$ structure function i.e. $dF_2 (x,Q^2)/d\ln Q^2$. We incorporated both the leading order(LO) and next-to-leading order (NLO) QCD contributions of the gluon-quark splitting kernels, in $dF_2 (x,Q^2)/d\ln Q^2$. Our result of $dF_2 (x,Q^2)/d\ln Q^2$ agrees reasonably well with  the experimental data recorded by HERA's H1 detector.   
\end{abstract}

\maketitle


\section{Introduction}
\label{intro}


Because of their universality in nature, parton distribution functions(PDFs) serve as very useful and basic tools in order to comprehend various standard model processes and in predictions of such proccesses at the accelerators. Consequently, various groups like the NNPDF\cite{ball2015parton}, CT\cite{ct14}, MMHT\cite{harland2018impact} and PDF4LHC\cite{pdf4lhc}  have been sincerely engaged in extracting and global fitting of PDFs. However, at very small momentum fraction x, among the partons, population of gluons become very high. This high population of gluons inside the hadrons lead to many nonlinear effects like saturation\cite{saturate111} and geometrical scalling\cite{geometrical_hera} of the hadronic cross sections.
\par So far, in description of PDFs, the Dokshitzer-Gribov-Levin-Altarelli-Parisi (DGLAP) equation\cite{ALTARELLI1977298}, which is a linear QCD evolution equation at the twist-2 level, has been extensively used with much phenomenological success. But, the solution of DGLAP equation towards very small-x predicts sharp growth of gluon densities. This would eventually violate (a) the unitarity of physical cross sections \cite{unitarity}; and (b) the Froissart bound of the hadronic cross sections at high energies \cite{froissart}. Therefore, the corrections of the higher order QCD effects, which suppress or shadow the unusual growth of parton densities, have become interesting topics of research in recent years.  
\par DGLAP equation was modified by incorporating correlation among the initial partons and considering various recombination processes by Gribov, Levin, Ryskin(GLR) in their pioneering work\cite{glr} at the twist-4 level; and later, Mueller and Qiu (MQ) performed perturbative calculation of the recombination probabilities in Double Leading Logarithmic Approximation (DLLA) \cite{mq1,mq2,mq3}, which enabled the GLR-MQ equation to be applied phenomenologically. The GLR-MQ equation sums up the gluon recombination diagrams using the Abramovsky-Gribov-Kancheli (AGK) cutting rules. This new evolution equation can serve as a tool to restore the unitarity as well as the Froissart bound. The basic difference between this nonlinear equation and the linear DGLAP equation is due to the presence of a shadowing term in the former. This shadowing term, which is quadractic in gluon density is coming from gluon recombinations inside the hadrons. In our previous work, we have studied extensively the gluon distribution functions by obtaining the solutions of GLR-MQ equation at leading order(LO)\cite{DEVEE2014571,devee2014nonlinear}, next-to-leading order(NLO)\cite{lalung2017nonlinear} and next-to-next-to-leading order(NNLO)\cite{LALUNG201929,phukan2017nnlo}. We observed the taming of gluon distribution function towards small-x as expected from the nonlinear GLR-MQ equation. However, there are also certain issues that seem to appear in the GLR-MQ equation: (a) the application of the AGK cutting rule in the GLR-MQ corrections breaks the evolution kernels\cite{zhu}; (b) the nonlinear term in the GLR-MQ equation violate the momentum conservation;\cite{zhu1993antishadowing} (c) the Double Leading Logarithmic approximation(DLLA) is valid only at small x and the GLR-MQ corrections cannot smoothly connect with the DGLAP equation \cite{zhu1}. These motivations led Zhu and his cooperators to derive a new QCD evolution equation including parton recombination in the leading logarithmic ($Q^2$) approximation (LL($Q^2$)A) using time ordered perturbation theory (TOPT) instead of the AGK cutting rules. This new evolution equation is popularly known as the GLR-MQ-ZRS equation \cite{zhu,zhu1,zhu2,zhu1993antishadowing,zhu2013emc,zhu2016chaotic}. The new evolution equation provides the following physical picture for the gluon recombination in a QCD evolution process: (a) the two-parton-to-two-parton ($2 \rightarrow 2$) amplitudes (Fig. 1a) give rise to the antiscreening effects; and (b) the interference amplitudes between the one-parton-to-two-parton ($1 \rightarrow 2$) (Fig. 1b) and the three-parton-to-two-parton ($3 \rightarrow 2$) (Fig. 1c) give rise to the screening effects, respectively.

\begin{figure}[t]
	\label{ fig7} 
	\begin{minipage}[b]{0.3\textwidth}
		\centering
		\includegraphics[width=.98\linewidth]{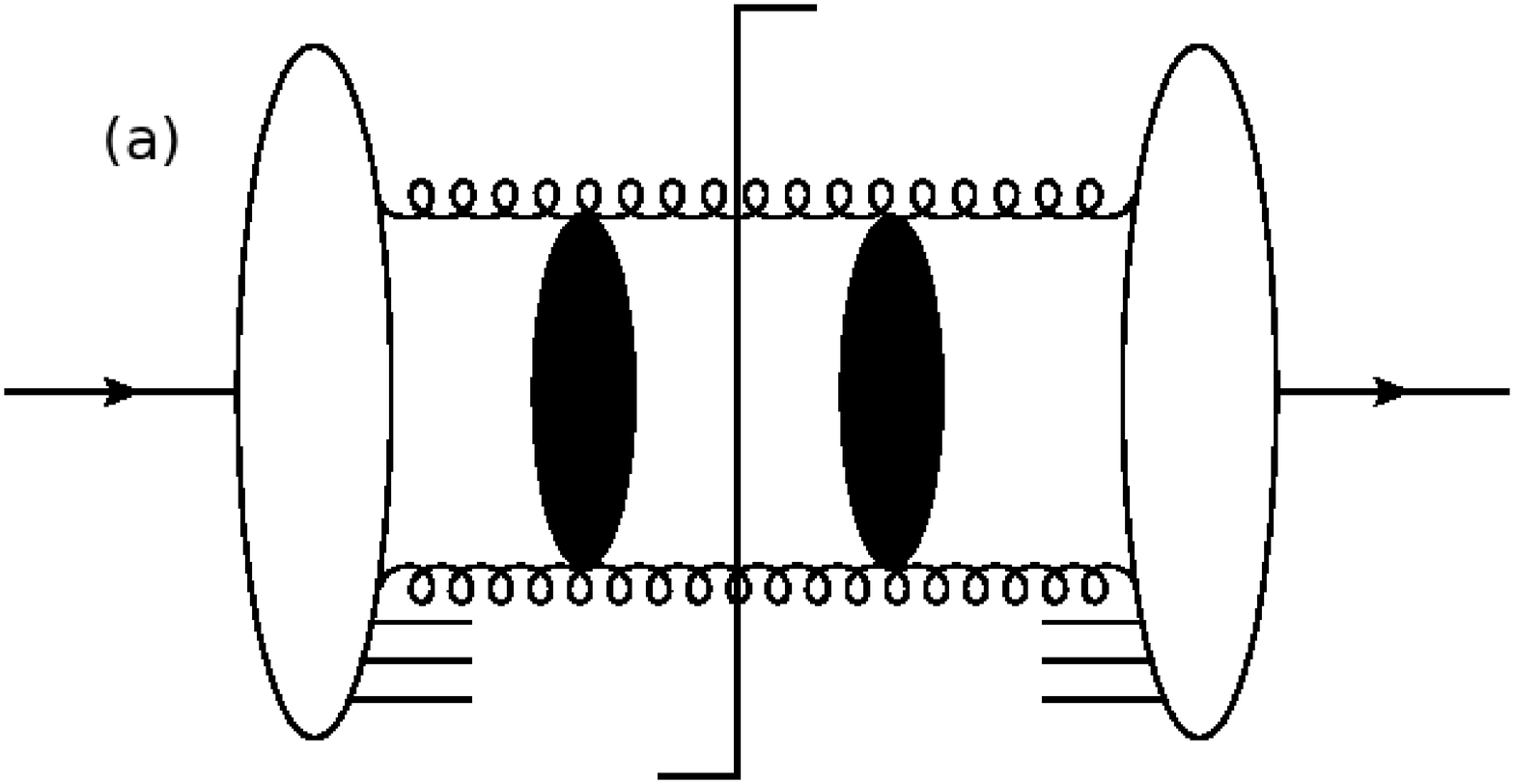} 
		
		
	\end{minipage}
	\hspace{0.25cm}
	\begin{minipage}[b]{0.3\textwidth}
		\centering
		\includegraphics[width=.98\linewidth]{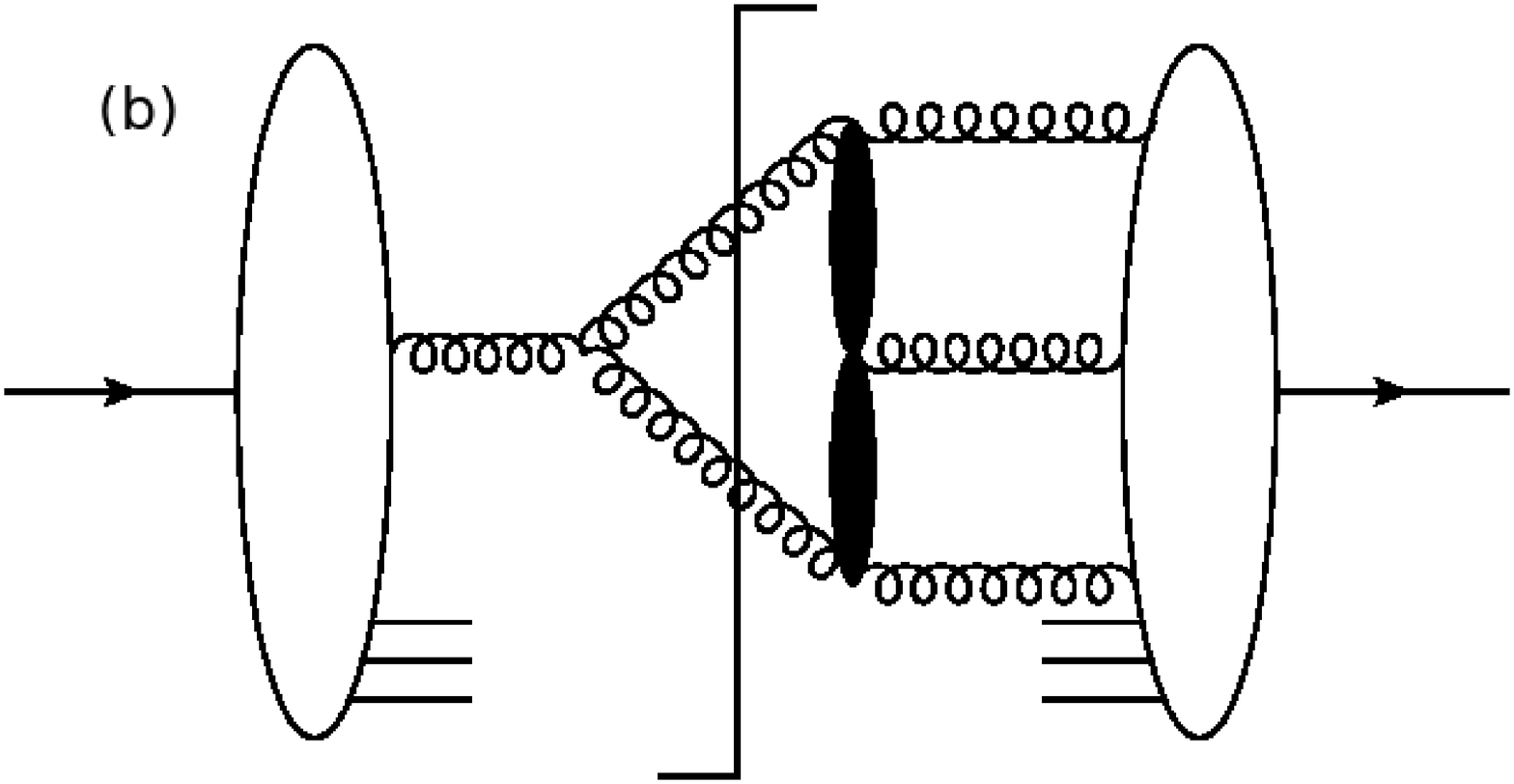} 
	\end{minipage}
	\hspace{0.25cm}
	\begin{minipage}[b]{0.3\textwidth}
		\centering
		\includegraphics[width=.98\linewidth]{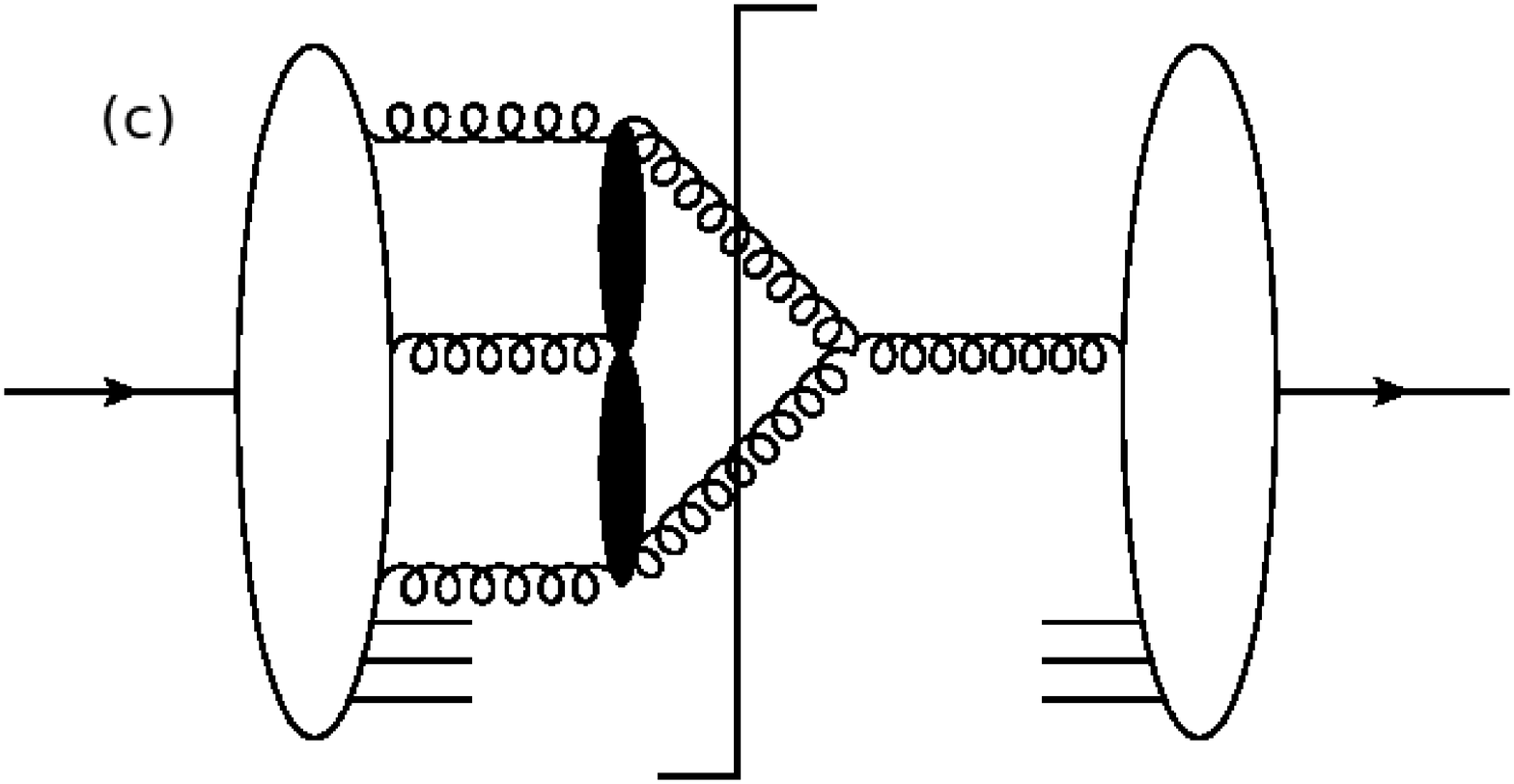} 
	\end{minipage}
	
	\caption{\small The Feynman diagrams contributing to the ZRS corrections with gluon
		recombination\cite{zhu}. The shaded part implies the correlation of gluons at short-distance.}
\end{figure}

\par  Also, in the high energy description of QCD, the triple pomeron vertex(TPV) has attracted significant attenttion in recent years \cite{IANCU2005253,Braun2006,Bartels2018} and its contribution is studied in perturbative QCD. It was originally derived from the $2 \rightarrow 4$ transition vertex in QCD reggeon field theory.  In the momentum space analysis of TPV in QCD, the authors in the work \cite{Bartels2008} have shown that after angular averaging, the TPV does not contribute to the collinear limit in the GLR-MQ nonlinear equation. Hence, their analysis doesnot agree with the form of the nonlinear term in the GLR-MQ equation. On the other hand, their analysis agrees well with Balitsky-Kovchegov (BK) equation \cite{BALITSKY199699,bk1,bk2,BALITSKY2001235,bk3,Kutak2005} in which the nonlinearity is given by TPV. The GLR-MQ-ZRS equation, however, was derived from the $2 \rightarrow 1$, $2 \rightarrow 2$ and $2 \rightarrow 3$ interference amplitudes at the leading order level, that is, at $1/(RQ)^2$ and $\alpha_s ^2$. GLR approach invloves summing up three kinds of diagrams: cutting two-ladders, one-
ladder and zero-ladder, respectively \cite{zhu}. While, in both GLR and ZRS approach the two ladder diagrams are identical; but, the cut lines in the latter two diagrams(one ladder and zero-ladder), break the parton recombination in GLR. Time ordered perturbation theory (TOPT) is developed to establish the connections among different cut diagrams in ZRS work. It would be very interesting in the future to also include the contributions from TPV coming from $2 \rightarrow 4$ transitions in the ZRS version of nonlinear equations. Thus, this remains as a provision for future study in the field of pQCD.  
\par A major difference between the GLR-MQ equation and the ZRS version is that the momentum conservation is restored in the GLR-MQ-ZRS equation by the antishadowing corrections. Therefore, it is interesting to study how the predictions of GLR-MQ equation are changed due to this new correction term. The purpose of this work is to study the behaviour of gluon distributions in the photon virtuality $Q^2$ and momentum fraction $x$ at high gluon density using the GLR-MQ-ZRS equation. We obtain the solution of the GLR-MQ-ZRS equation by employing Regge like behaviour of gluons \cite{collins1977introduction} and perform phenomenological study of the solution in kinematic range of $5\,GeV^2\leq Q^2 \leq 100\, GeV^2$ and $10^{-6}\leq x \leq 10^{-2}$ respectively . It is to note that the Regge like behaviour of gluons are advocated well in the kinematic range of moderate $Q^2$ and small-x \cite{boroun2006solution,donnachie1998small,badalek,soffer}. Therefore, in this work we also check the validity of Regge ansatz on gluon distributions in the kinematic range of our consideration. We compare our results of gluon distribution functions with the results of various global PDF groups viz., CT14\cite{ct14}, NNPDF3.0\cite{ball2015parton} and PDF4LHC15\cite{pdf4lhc}. We have used the APFEL tool\cite{bertone2014apfel,carrazza2015apfel} with LHAPDF6 library\cite{buckley2015lhapdf6} to generate PDFs from the global groups. We have also compared our results of gluon distribution functons with the results obtained from the IMParton C++ package \cite{IMParton16,chen2014applications,zhu2016determination}. It is also known that at very small-x, the logarithmic derivative of proton's $F_2 (x,Q^2)$ structure function i.e. $dF_2 (x,Q^2)/d \ln Q^2$, has direct relationship with the gluon distribution function $G(x,Q^2)$ \cite{gotsman2001scaling,aid1995gluon}. Hence, we predict our results of $dF_2 (x,Q^2)/d \ln Q^2$ in $x$ and $Q^2$ using the solutions of GLR-MQ-ZRS equation and then compare our results with the HERA data \cite{adloff2001ep}. 
 \par The paper is organized as follows: In Sec. II we develope the formalism to solve the GLR-MQ-ZRS equation for obtaining the expressions for $G(x,Q^2)$ and $dF_2 (x,Q^2)/d \ln Q^2$. In. Sec. III we discuss our results and finally, we conclude in Sec. IV.

\section{Formalism}

The main difference between GLR-MQ and GLR-MQ-ZRS equation is that in the former case only the shadowing term is present while both the shadowing as well as antishadowing terms are incorporated in the latter. The ZRS corrections to the of Altarelli Parisi equation with recombination functions in DLLA is given by\cite{zhu} 

\begin{equation}
\begin{split}
\frac{dG(x,Q^2)}{d\ln{Q^2}} = \frac{\alpha_s (Q^2) N_c}{\pi}\int_{x}^{1} \frac{dy}{y} G(y,Q^2) +& \frac{9}{2\pi}\cdot \frac{\alpha_s ^2 (Q^2)}{R^2 Q^2}\cdot \frac{N_c ^2}{N_c ^2 -1}\int_{x/2}^{1/2} \frac{dy}{y}G^2(y,Q^2) \\
-&\frac{9}{\pi}\cdot \frac{\alpha_s ^2 (Q^2)}{R^2 Q^2}\cdot \frac{N_c ^2}{N_c ^2 -1}\int_{x}^{1/2} \frac{dy}{y}G^2(y,Q^2),
\end{split}
\end{equation}

Here, the representation for the gluon distribution $G(x,Q^2) =xg(x,Q^2)$is used, where $g(x,Q^2)$ is the gluon density per transverse area inside the hadron. The first term on the R.H.S. of the above equation is the usual DGLAP equation at Double Logarithmic Approximation (DLA). The second and the third term represents the antishadowing and shadowing contributions from the correlative gluons inside the hadrons respectively. $N_c$ is the number of color charges and $R$ represents the correlative radius of the gluons inside the hadrons. Unlike the GLR-MQ equation where the strong growth of gluons generated by the linear term was tamed down by the shadowing term; in GLR-MQ-ZRS equation, it is the net collective effect due to both the shadowing and antishadowing terms, because of which the unusual growth of the gluons get tamed down. Another important difference between GLR-MQ and GLR-MQ-ZRS equation, as can be seen in Eq. (1), is that the positive antishadowing effects are separated from the negative shadowing effects because of the fact that they have different kinematical domains in x.

 The size of the nonlinear terms depends on the value of the correlation length $R$. If the gluons are populated across the hadron(say proton) then $R \approx 5\,GeV^{-1}$, and if the gluons have hotspot like structure then $R \approx 2\,GeV^{-1}$.

\par The DGLAP equation at DLA by considering all the splitting functions and strong coupling constant $\alpha_s (Q^2)$ at leading orders (LO) is given by
\begin{equation}
\begin{split}
\frac{d G(x,Q^2)}{d \ln{Q^2}}\bigg|_{DGLAP} = &\frac{3\alpha_{s}(Q^2)}{\pi}\bigg[\bigg\{ \frac{11}{12}-\frac{N_f}{18}+\ln(1-x) \bigg\} G(x,Q^2) \\
&  +\int_{x}^{1} dy \cdot\bigg\{\frac{y G(\frac{x}{y},Q^2)-G(x,Q^2)}{1-y}\bigg\}   \bigg] \\ &+\frac{3\alpha_{s}(Q^2)}{\pi}\int_{x}^{1} dy \bigg\{y (1-y )+\frac{1-y}{y}\bigg\} G({x}/{y},Q^2),
\end{split}
\end{equation} 
where  $\alpha_s ^{LO}/4\pi = 1/\beta_0 \ln{(Q^2/\Lambda^2)}$, $\Lambda$ is the QCD cutoff parameter and $\beta_0 = 11 - (2/3) N_f$. The number of flavors $N_f$ is taken to be 4 in our calculations.

 Here, we have ignored the quark gluon emission diagrams due to their little importance in the gluon rich small-x region. The kinematic range in which we want to study the behaviour of gluons is $10^{-6} \leq x < 10^{-2}$ and $5 \, GeV^2 \leq Q^2 \leq 100 \, GeV^2$ respectively.

 At small-x, the behaviour of structure functions is well explained in terms of Regge-like behaviour . The behaviour of structure functions at small-x for the fixed photon virtuality  $Q^2$ reflects the high-energy behaviour in which the total cross section shows a power like behaviour in terms of the total CM energy squared $s^2$, where $s^2=Q^2(1/x-1)$ \cite{Kwiecinski}. 
 The Regge pole exchange picture \cite{collins1977introduction} would therefore appear to be quite appropriate for the theoretical description of this behaviour. For the sea-quark and antiquark distributions, the small-x Regge behaviour is given by the power law $q_{sea}(x)\sim x^{-\alpha_P}$ corresponding to a pomeron exchange with an intercept of $\alpha_P= 1$. Whereas valence quark distributions are governed by the reggeon exchange with an intercept of $\alpha_R=0.5$. It is to note that only at the moderate $Q^2$, the x dependence of the parton densities is assumed and the leading order calculations in $\ln(1/x)$ with fixed $\alpha_S$ predict a steep power-law behaviour of $xg(x,Q^2)\sim x^{−\lambda_G}$, where $\lambda_G= (3\alpha_{s}/\pi)4\ln2\approx 0.5$ for $\alpha_{s}\approx0.2$, as appropriate for $Q^2\sim4GeV^2$.

 \par In the Regge inspired model developed by Donnachie and Landshoff (DL), the HERA data could be fitted very well by considering  
 the exchange of both the soft and hard pomerons contributing to the amplitude and at small-x the gluon distribution function is dominated by hard pomeron exchange alone \cite{donnachie1998small,donachi,donnachie2002proton}. In the DL model, the simplest fit to the small-x data corresponded to $F_2(x,Q^2) =A(Q^2)x^{-\epsilon_0}$,with $\epsilon_0= 0.437$ [43-45]. They have also shown that the result of integration of the differntial equation ${\partial F_2(x,Q^2)}/{\partial \ln Q^2} \sim P_{qq} \otimes G(x,Q^2)$ at small x, and the gluon distribution function $G(x,Q^2)$ is described by the exchange of a hard pomeron.

 So, Regge theory provides a highly ingenous parametrization of all total cross sections and is supposed to be applicable for very small-x and whatever be the value of $Q^2$ as far as the quantity $W^2 = 1/Q^2 (1/x -1 )$ is greater than all other variables. Models based upon this idea have been successful in describing the DIS cross-section when $x$ is small enough $(x < 0.01)$, whatever be the value of $Q^2$ \cite{bartel_regge,Martin_regge}.  Within the kinematic range in which we want to perform our study, behaviour of gluon distribution function can be well explained in terms of the Regge ansatz\cite{donnachie1998small,badalek,soffer}. For this reason Regge pole exchange picture sounds convenient in theoretical description of behvaiour of structure functions. Therefore, we will also use a simple form of Regge behaviour for $G(x,Q^2)$ as given below:  
\begin{equation}
G(x,Q^2)=x^{-\lambda_G}f(Q^2),
\end{equation}

\noindent where $\lambda_G$ is  the Regge intercept. $\lambda_G$ is related to the QCD coupling constant $\alpha_s$ via the relation $\lambda_G = (4N_c \alpha_s/\pi)\ln{2}$\cite{kwiecinski1993qcd}.

This form of Regge behaviour is supported well by many authors in their work\cite{boroun2006solution,donnachie1998small,badalek,soffer}. One of the important features of the Regge theory is that, at small-x the behaviour of both the gluons and sea quarks are controlled by the same singularity factor in the complex angular momentum plane.

Now, in order to simplify Eq. (1), we consider following assumtions
\begin{itemize}
	\item The value of gluon distribution function $G(x,Q^2)$ in the region of $x$ asymptotically decreasing towards $0$, is much greater than the value of $G(x,Q^2)$ at Bjorken $x= 1/2$.
	\item On probing the large-momentum part of the proton structure ($x\approx1$), the valence quarks dominate. Hence,  $G(x,Q^2)\approx 0$ at Bjorken $x \rightarrow 1$.  
\end{itemize}

\par Applying all the assumptions mentioned above and using Eqs. (2) and (3) in Eq. (1), and applying Taylor series expansion on $G(x,Q^2)$ at suitable points in the basis of $x$ at small-x, Eq. (1) can then be simplified into the following form

\begin{equation}
\begin{split}
\frac{d G(x,Q^2)}{d \ln{Q^2}}=\frac{d G(x,Q^2)}{d \ln{Q^2}}\bigg|_{DGLAP}+\frac{81}{32\pi \lambda_G}\cdot \frac{\alpha_s ^2 (Q^2)}{R^2 Q^2}&\frac{G^2(x,Q^2)}{\tau^2_{x/2}} \\
-&\frac{81}{16\pi \lambda_G}\cdot \frac{\alpha_s ^2 (Q^2)}{R^2 Q^2}G^2(x,Q^2),
\end{split}
\end{equation}

where $\tau_{x/2}= 1- \lambda_G/2 + \lambda_G ^2/2$. This factor $\tau_{x/2}$ is coming from the Taylor series expansion of $G(x,Q^2)$ at Bjorken $x/2$.


To obtain the solution of Eq. (4), first we keep the strong coupling constant $\alpha_s$ fixed. Here, we assume no dependency of $\alpha_s$ on $Q^2$ at all. Now, it is convenient to use a new variable given as $t = \ln{(Q^2/\Lambda^2)}$, $\Lambda$ is the cutoff parameter. In terms of this new variable $t$ and on further simplications, Eq. (4) can conveniently be expressed as 
\begin{equation}
\frac{d G(x, t)}{dt}= P(x)\cdot G(x,t) + \frac{\chi - \psi}{ e^t}G^2 (x,t),
\end{equation}
The functions involved are 
\begin{equation*}
\begin{split}
&P(x)=\frac{4\pi}{\beta_0}\left(\frac{11}{12}-\frac{N_f}{18}+\ln{(1-x)}+\int_x^1 dy(\frac{y^{\lambda_G +1}-1}{1-y}+ ( y(1-y) + \frac{1-y}{y} ) )\right) \\
&\chi= \frac{81\pi}{2 \lambda_G R^2 \beta_0 ^2 \Lambda^2 \tau^2_{x/2} } \,\,\,\text{and}\,\,\, \psi = \frac{81\pi}{ \lambda_G R^2 \beta_0 ^2 \Lambda^2 }
\end{split}
\end{equation*}

The function $P(x)$ is coming from the leading order Alaterilli Parisi kernels of the gluon-gluon emssion $(P_{gg})$ \cite{furmanski1980singlet}. $\chi$ and $\psi$ are coming from the shadowing and antishadowing corrections respectively.  Eq. (5) looks similar to the Bernoulli's nonlinear differential equation which can be solved easily. The analytical solution has the following form
\begin{equation}
 G(x,t)= \frac{P(x)-1}{C(P(x)-1)e^{-P(x) t}-(\chi -\psi) e^{-t_0}},
\end{equation}

Where C is the constant of integration. C is to be determined using suitable initial conditions. When both the shadowing and antishadowing terms balance each other, $(\chi -\psi) $ becomes zero and the solution is of the form $C\cdot exp[-P(x)\ln t]$. The Double Leading Logarithmic (DLLA) solution of DGLAP equation reads as $G(x,t)\propto exp[C \ln t \ln (1/x)]^{1/2} $. It is to note that Regge behaviour is not in agreement with DLLA, altough the range in which $x$ is very small and $Q^2$ is not too large, is actually the Regge regime. From Eq. (6),  the $x$ and $t$ evolutions of $G(x,t)$ can be determined separately. We can easily return back to our original variables $x$ and $Q^2$ by using the relation $t=\ln (Q^2/\Lambda ^2)$. 
\par Now, to obtain the $x$ evolution of $G(x,Q^2)$, we will consider a suitable input $G(x_0,Q^2)$ at larger value of $x_0$ for a particular $Q^2$. Then, from Eq. (6) we have 

\begin{equation}
	 G(x_0,t)= {\left(P(x_0)-1\right)}\times\left({C(P(x_0)-1)e^{-P(x_0) t}-(\chi -\psi) e^{-t_0}}\right)^{-1},
\end{equation}
for which we obtain the value of the constant C as

\begin{equation}
	C = \left({\left(P(x_0)-1 \right) + G(x_0,t)(\chi -\psi) e^{-t_0}}\right)\times \left({ G(x_0,t)(P(x_0)-1)e^{-P(x_0) t}}\right)^{-1}
\end{equation}
Finally, using Eq. (8) in Eq. (6) we obtain the $x$ evolution(for $x \leq x_0$) as
\begin{equation}
\begin{split}
&G(x,Q^2) = \\&G(x_0,Q^2) (P(x)-1)(P(x_0)-1)e^{-P(x_0)\ln(Q^2/\Lambda ^2)} \times \Bigg( (P(x)-1)(P(x_0)-1)e^{-P(x)\ln(Q^2/\Lambda ^2)} \\ &+ G(x_0,Q^2)(\chi -\psi)  e^{-\ln (Q^2/\Lambda ^2)}\left(e^{-P(x)\ln(Q^2/\Lambda^2)}(P(x)-1)-e^{-P(x_0)\ln(Q^2/\Lambda^2)}(P(x_0)-1)\right)\Bigg)^{-1}
\end{split}
\end{equation}
\par To obtain the $Q^2$ evolution of $G(x,Q^2)$, we take the input $G(x,Q_0 ^2)$ at a lower value of $Q_0 ^2$ for a particular Bjorken $x$. The $Q^2$ evolution (for $Q^2\geq Q_0 ^2$) is given by

\begin{equation}
\begin{split}
&G(x,Q^2)= G(x,Q_0 ^2)(P(x)-1)e^{-P(x)\ln (Q_0 ^2 /\Lambda ^2)} \times \Bigg((P(x)-1)\ln (Q ^2 /\Lambda ^2)\\ &+ G(x,Q_0 ^2)(\chi -\psi) \left(  e^{-\ln (Q_0 ^2 /\Lambda ^2)- P(x)\ln (Q ^2 /\Lambda ^2)} - e^{-\ln (Q ^2 /\Lambda ^2)- P(x)\ln (Q_0 ^2 /\Lambda ^2)} \right) \Bigg)^{-1}
\end{split}
\end{equation}
 
After working out the $x$ and $Q^2$ evolutions separately, we can merge the Eqs. (9) and (10) together so that we can obtain both the $x$ and $Q^2$ evolutions from a single equation. We observe that in Eq. (9), the $x$ evolution equation of $G(x,Q^2)$ consists of the input $G(x_0,Q^2)$ at $(x_0, Q^2)$ and is multiplied with the $x$ evolution part of the equation. Similarly, the $Q^2$ evolution of $G(x,Q^2)$ consists of the input $G(x,Q_0 ^2)$ at $(x,Q_0 ^2)$ and is multiplied with the $Q^2$ evolution part of the equation. We first take the $x$ evolution of $G(x,Q^2)$,for which the input $G(x_0,Q^2)$  can obtained from another input point at $(x_0,Q_0 ^2)$ by using the $Q^2$ evolution given in Eq. (10). Thus we can merge both the $x$ and $Q^2$ evolutions together in a single equation. The merger of these two evolutions is given by

\begin{equation}
G(x,Q^2)= G(x_0, Q_0 ^2)\times [ \text{x evolving part}]_{Q^2}\times [ \text{$Q^2$ evolving part}]_{x_0}
\end{equation}


In the second case, we consider the leading order(LO) of the running coupling constant $\alpha_{s} (Q^2)$, Eq. (4) can be simplified into the following form

\begin{equation}
\begin{split}
\frac{d G(x, t)}{dt}= \frac{P(x)}{t}G(x,t) + \frac{\chi - \psi}{t^2 e^t}G^2 (x,t),
\end{split}
\end{equation}

\par Now, we are interested in obtaining the $x$ and $t$ (or $Q^2$)-evolution of gluon distribution function $G(x,Q^2)$. Eq. (12) can also be simplified into a form of Bernoulli's nonlinear differential equation, which then can be solved using integrating factor. We follow the same procedure as in the case of fixed $\alpha_s$  to obatin $x$ and $Q^2$ evolution of $G(x,Q^2)$. The solutions of Eq. (12) are expressed in terms of the Incomplete Gamma functions, which can be evaluated numerically. Therefore, the results in this case are expressed in semi-analytical form.   
We use the following definition
$$\Gamma [s,x] = \int_x^{\infty}x^{s-1}e^{-x}dx$$
The $x$ evolution of $G(x,Q^2)$ using an input at $x_0$ (for $x \leq x_0$) is given by 
\begin{equation}
\begin{split}
&G(x,Q^2)= G(x_0,Q ^2 ) e^{P(x)\ln{(\ln{(Q^2/\Lambda^2)})}}\times \Bigg( e^{P(x_0)\ln{(\ln{(Q^2/\Lambda^2)})}} \\ &+ G(x_0,Q^2 )(\chi - \psi)\left( \Gamma[ P(x)-1,\ln{(Q^2/\Lambda^2)}] -\Gamma[ P(x_0)-1, \ln{(Q^2/\Lambda^2)}] \right) \Bigg)^{-1}
\end{split}
\end{equation}

Similarly, the $Q^2$ evolution of $G(x,Q^2)$ using an input at $Q_0 ^2$ (for $Q^2 \geq Q_0 ^2$) is given by  
	\begin{equation}
\begin{split}
&G(x,Q^2)= G(x,Q_0 ^2 ) e^{P(x)\ln{(\ln{(Q^2/\Lambda^2)})}}\times \Bigg( e^{P(x)\ln{(\ln{(Q_0^2/\Lambda^2)})}} \\ &+ G(x,Q_0 ^2 )(\chi - \psi)\left(\Gamma[ P(x)-1,\ln{(Q^2/\Lambda^2)}] -\Gamma[ P(x)-1, \ln{(Q_0^2/\Lambda^2)}]\right) \Bigg)^{-1}
\end{split}
\end{equation} 	

Eqs. (13) and (14) can be merged together using Eq. (11) as mentioned in the previous section.
\par Now, it is useful to study the contributions coming from the antishadowing and shadowing effects on the gluon distribution function. If the antishadowing effect is switched off, then the GLR-MQ-ZRS turns into the GLR-MQ equation. We have already solved the GLR-MQ equation using Regge ansatz and performed phenomenological study of $G(x,Q^2)$ in our previous work \cite{DEVEE2014571,devee2014nonlinear}. To study the antishadowing contribution, we define a parameter $R_G$ such that it gives us the ratio of $G(x,Q^2)$ from the solution of GLR-MQ equation to $G(x,Q^2)$ from the solution of GLR-MQ-ZRS equation. This parameter is useful to quantify the contribution of antishadowing corrections with respect to the shadowing corrections on the gluons. Thus, we define

\begin{equation}
	R_G(x,Q^2) = \frac{G^{glr-mq}(x,Q^2)}{G^{glr-mq-zrs}(x,Q^2)}
\end{equation}

\begin{figure}[t]
	\centering
	\includegraphics[width=0.9\textwidth]{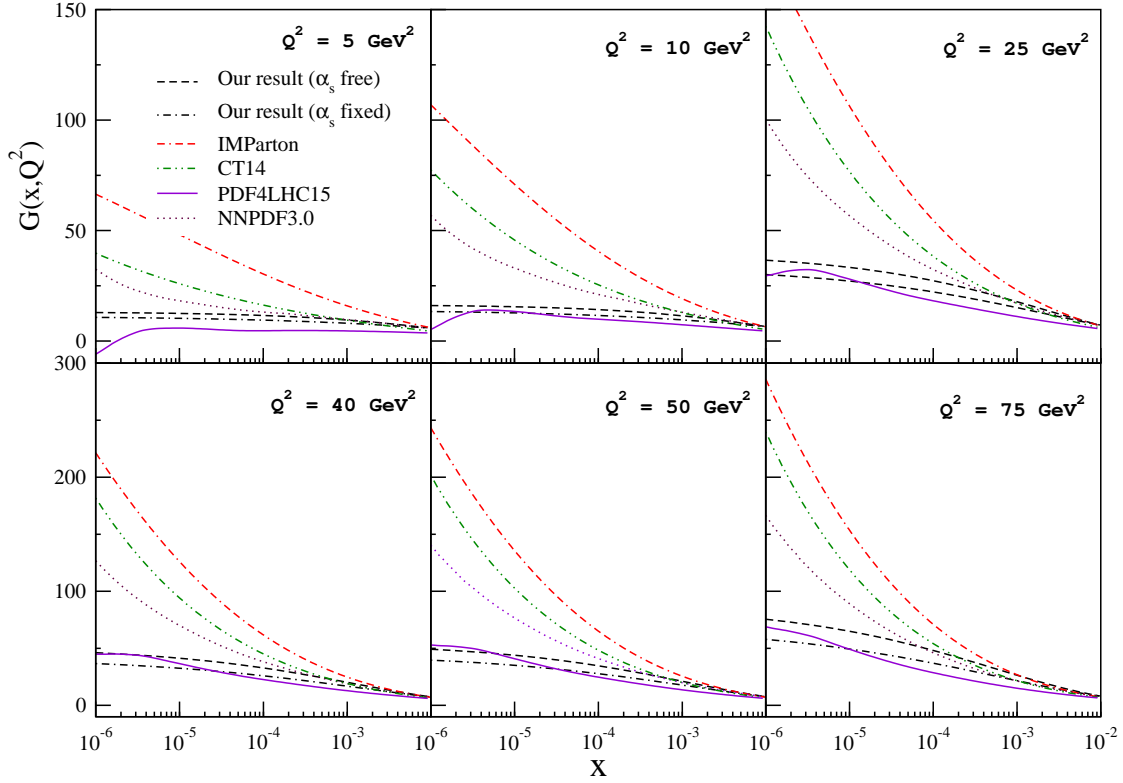}
	\caption{\small $x$ evolution of $G(x,Q^2)$ for fixed values of photon virtuality or four momentum transfer squared ($Q^2$). The correlation radius $R$ is taken to be $5\, GeV^{-1}$. The dashed line represents our result with the $Q^2$ dependency of $\alpha_{s}$, whereas the dashed dot line represents our result with $\alpha_{s}$ constant($\approx 0.2$). Our results are compared with NNPDF3.0, CT14, PDF4LHC15 and IMParton.   }
	\label{fig:1}       
\end{figure}

\begin{figure}[t]
	\centering
	\includegraphics[width=0.9\textwidth]{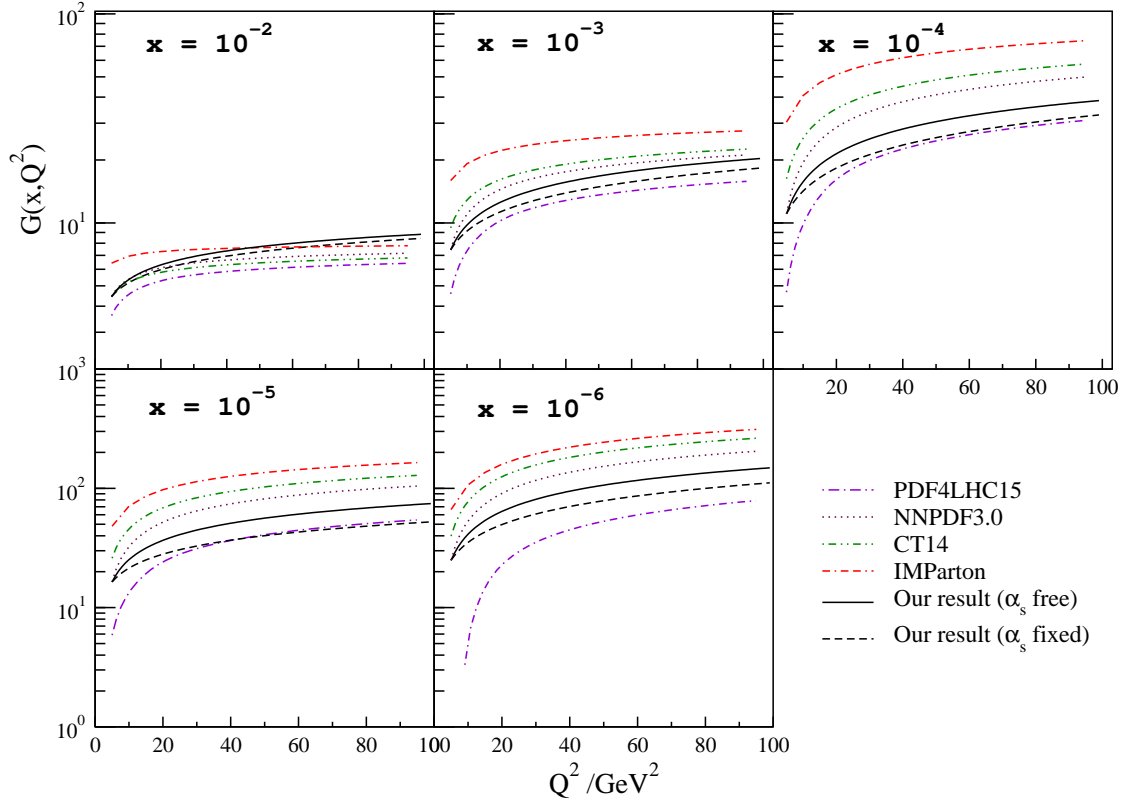}
	\caption{\small $Q^2$ evolution of $G(x,Q^2)$ for fixed values of momentum fraction or Bjorken $x$. The correlation radius $R$ is taken to be $5\, GeV^{-1}$. The solid line represents our result with the $Q^2$ dependency of $\alpha_{s}$, whereas the dashed line represents our result with $\alpha_{s}$ constant($\approx 0.2$).  }
	\label{fig:1}       
\end{figure}


In order to compare our results with the experimental data, we investigate the net effect due to both the shadowing and antishadowing corrections to the evolution of the singlet quark distribution with respect to GLR-MQ-ZRS
equations. It can be shown using the DGLAP evolution equation that the $Q^2$ logarithmic slope of proton's structure function $F_2$ at small-x, in leading order, is directly proportional to the gluon distribution function $G(x,Q^2)$. Several methods have been reported in relating the scaling violations of $F_2$ with the gluon
density at small-x \cite{logariBoroun,aid1995gluon,prytz1993approximate}. All methods are based on an approximate relation,
using the fact that at small-x, quark densities can be neglected and that the nonsinglet contribution $F_2 ^{NS}$ to the overall $F_2$ structure function can be ignored safely. In particular, the solution of the DGLAP equation predicts sharp rise of the structure function almost increasing with the powers of x towards small-x. This steep rise of $F_2 (x, Q^2 )$ towards small-x was  observed at HERA. Since, the region of small-x, is the region where gluons get overpopulated inside the proton; thus it is interesting to study the behaviour of the logarithmic growth of $F_2$ with respect to the overall growth of the gluon density. 

\par The $F_2$ structure function of proton is the sum of the contributions coming from both the singlet $F^S _2$ and nonsinglet $F^{NS} _2$ structure functions, 

\begin{equation}
F_2= \frac{5}{18}F^S _2+ \frac{3}{18}F^{NS} _2
\end{equation}

The nonsinglet part of Eq. (16) can be neglected in view of the overpopulation of gluons among the partons in the small-x region. Therefore, at small-x, $F_2$ is dictated by the singlet contributions alone. The logarithmic derivative of $F_2$  then reads as,

\begin{equation}
\begin{split}
\frac{dF_2(x,Q^2)}{d\ln{Q^2}} = P^{AP} _{qg}\otimes G(x,Q^2) +& \frac{9}{2\pi}\cdot \frac{\alpha_s ^2 (Q^2)}{R^2 Q^2}\cdot \frac{N_c ^2}{N_c ^2 -1}\int_{x/2}^{1/2} \frac{dy}{y}G^2(y,Q^2) \\
-&\frac{9}{\pi}\cdot \frac{\alpha_s ^2 (Q^2)}{R^2 Q^2}\cdot \frac{N_c ^2}{N_c ^2 -1}\int_{x}^{1/2} \frac{dy}{y}G^2(y,Q^2),
\end{split}
\end{equation}

Where, the convolution $\otimes$ represents the prescription $f(x)\otimes g(x)=\int_{x}^{1}dy/y f(y)g(x/y)$ and  $P_{qg}$ is the AP kernel for gluon to quark-antiquark emissions \cite{vogt2004three,furmanski1980singlet}. The quark-quark splitting function $P_{qq}$ are ignored, since maximum contributions are due to the overpopulated gluons only. $P_{qg}$ can be expanded perturbatively in powers of $\alpha_{s}$ as

\begin{equation}
P^{AP}_{qg} (x)= P^{(0)}_{qg} (x) + \frac{\alpha_s}{2\pi}P^{(1)}_{qg} (x)+\cdots ,
\end{equation}
Where $P^{(0)}_{qg} (x)$ and $P^{(0)}_{qg} (x)$ are the leading order(LO) and next-to-leading order (NLO) contributions to the splitting function rspectively \cite{vogt2004three}. The expressions of these splitting functions are given by

\begin{equation*}
\begin{split}
&P^{(0)} _{qg} (x)= 2T_f P_{qg} (x), \\
&P^{(1)}_{qg} (x)= C_F T_f \bigg\{ 4-9x-(1-4x)\ln x -(1-2x)\ln^2 x + 4 \ln(1-x) \\ &+\left(2\ln^2 (\frac{1-x}{x})-4\ln (\frac{1-x}{x})- 2/3 \pi ^2 +10\right)P_{qg} (x) \bigg\}+ N_c T_f \bigg\{ \frac{182}{9} + \frac{14x}{9}+\frac{40}{9x}\\&+\left(\frac{136x}{3}-4\ln(1-x)\right)  -(2+8x)\ln^2 x + \bigg( -\ln^2 x + \frac{44}{3}\ln x - 2\ln ^2 (1-x)+ 4\ln(1-x)\\&+\frac{\pi ^2}{3}-\frac{218}{9}\bigg)P_{qg}(x) + 2P_{qg}(-x)S_2 (x) \bigg\} ,\\
\end{split}
\end{equation*}

\begin{equation}
\begin{split}
&S_2 (x)= \int_{\frac{x}{1+x}}^{\frac{1}{1+x}}\frac{dz}{z}\ln (\frac{1-z}{z})\xrightarrow[x]{\text{small}} \frac{1}{2}\ln ^2 x - \frac{\pi ^2}{6} + O(x),\\
&P_{qg} (x)=  x^2 + (1-x)^2 \, \text{and}\,C_F = 4/3,\,N_c=3,\, T_f=N_f /2.
\end{split}
\end{equation}  

Using the Regge like behaviour of gluons, Eq. (17) can be written as
\begin{equation}
\begin{split}
\frac{dF_2(x,Q^2)}{d\ln{Q^2}} = \frac{5\alpha_{s}}{9\pi}T(\lambda_G) G(x,Q^2)+& \frac{9}{2\pi}\cdot \frac{\alpha_s ^2 (Q^2)}{R^2 Q^2}\cdot \frac{N_c ^2}{N_c ^2 -1}\int_{x/2}^{1/2} \frac{dy}{y}G^2(y,Q^2) \\
-&\frac{9}{\pi}\cdot \frac{\alpha_s ^2 (Q^2)}{R^2 Q^2}\cdot \frac{N_c ^2}{N_c ^2 -1}\int_{x}^{1/2} \frac{dy}{y}G^2(y,Q^2),
\end{split}
\end{equation}
\begin{equation*}
T(\lambda_G)= \int_{x}^{1}y^{\lambda_G}P^{AP}_{qg} (y)dy.
\end{equation*}

Finally, using Eq. (20) and from the solutions of $G(x,Q^2)$ given in Eq. (13) and (14), we can now obatin the logarithmic derivative of proton's $F_2 $ structure function.

\begin{figure}[t]
	
	\begin{minipage}[b]{0.46\textwidth}
		\centering
		\includegraphics[width=\linewidth]{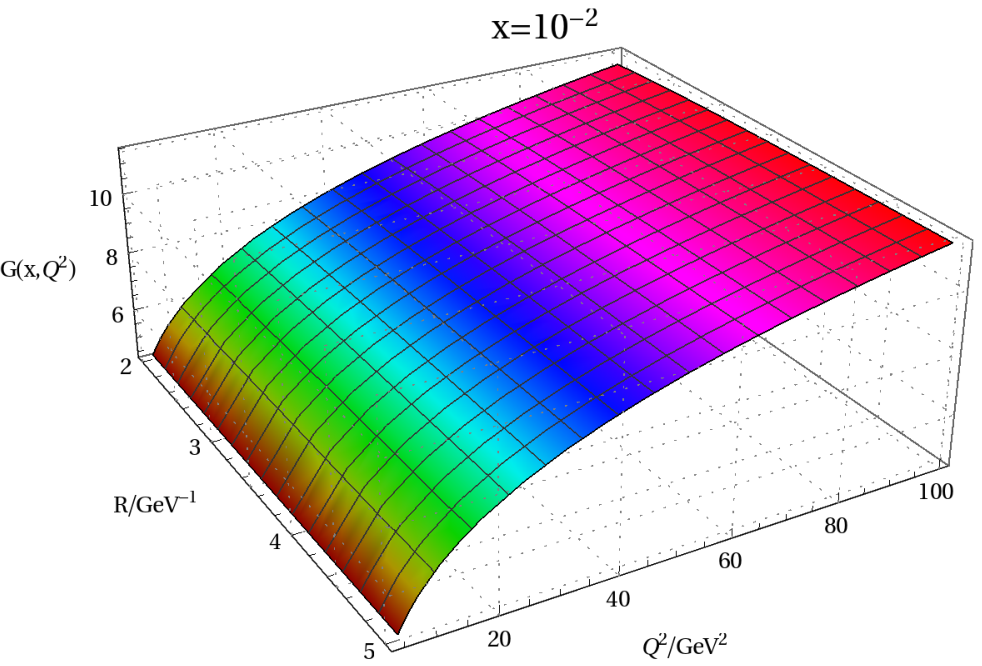} 
		
		
	\end{minipage}
	\hspace{1.0cm}
	\begin{minipage}[b]{0.46\textwidth}
		\centering
		\includegraphics[width=\linewidth]{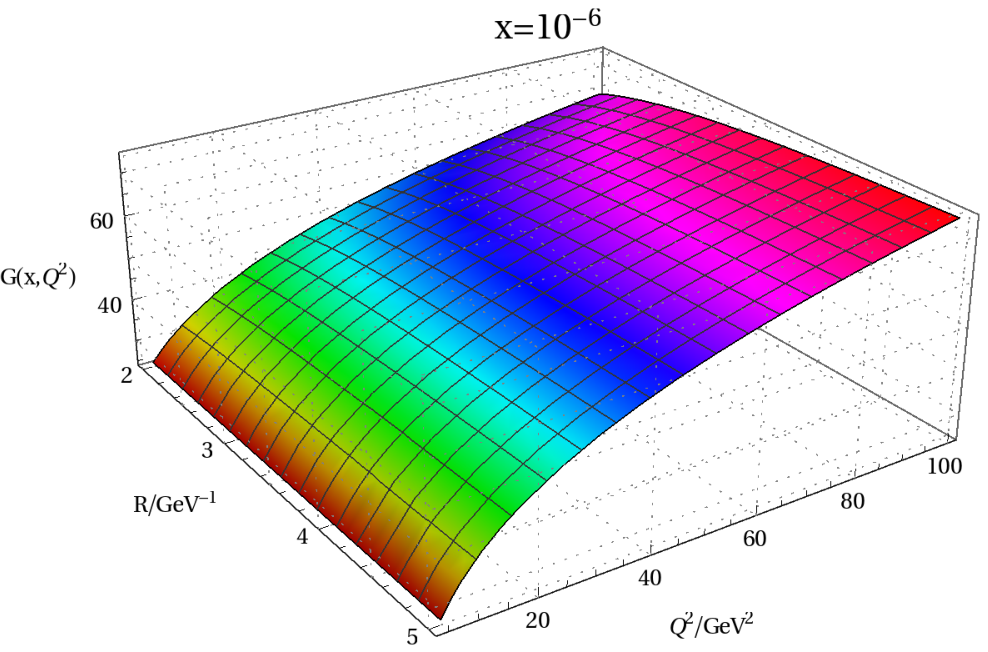} 
	\end{minipage}

	\begin{minipage}[b]{0.46\textwidth}
		\centering
		\includegraphics[width=\linewidth]{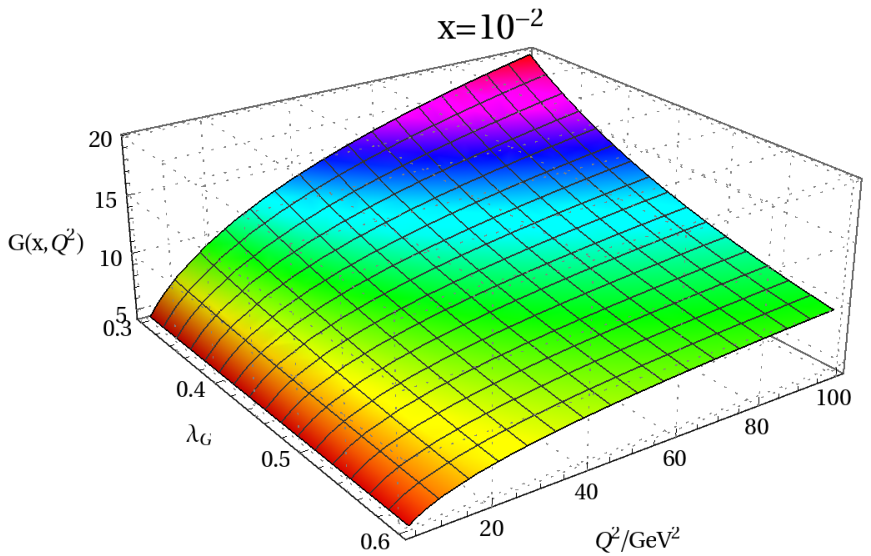} 

	\end{minipage}
	\hspace{1.0cm}
	\begin{minipage}[b]{0.46\textwidth}
		\centering
		\includegraphics[width=\linewidth]{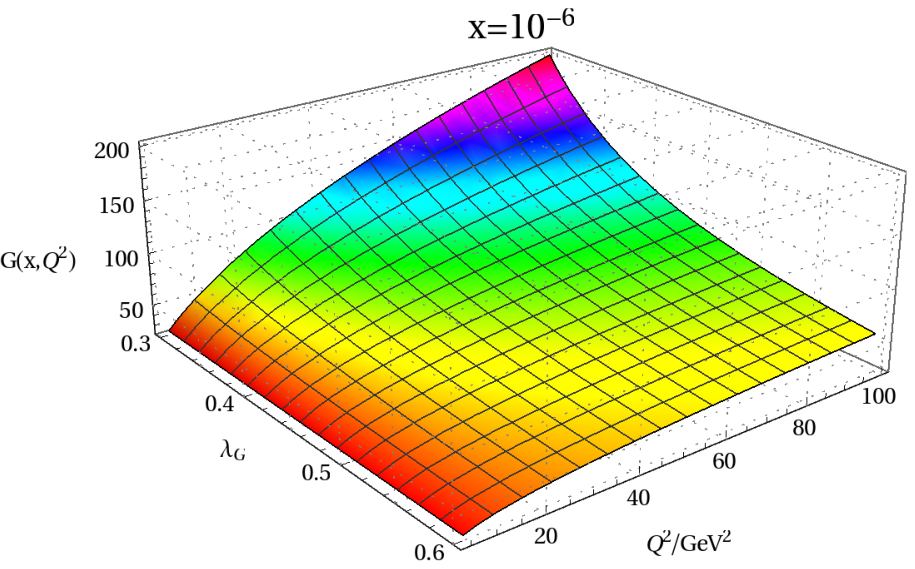} 
		
		
	\end{minipage}
	\caption{\small The sensitivities of the Regge intercept $\lambda_G$ and correlation radius $R$ on $G(x,Q^2)$ are shown with respect to $Q^2$ at $x=10^{-2}$ and $10^{-6}$ respectively. } 
\end{figure}

\begin{figure}[h]
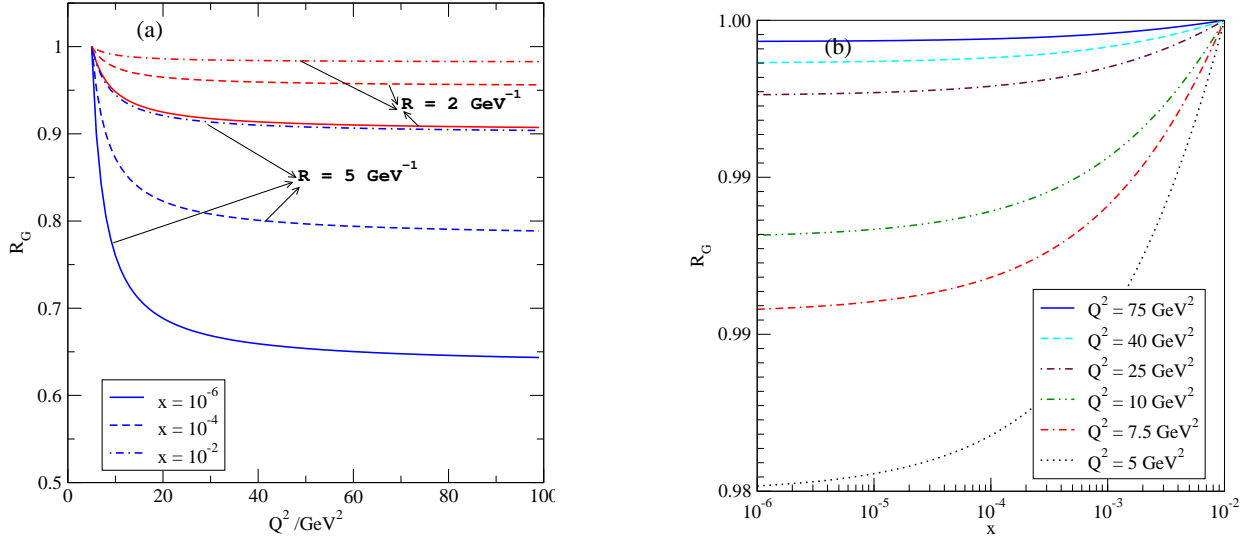

	\label{ fig7} 
	\begin{minipage}[b]{0.45\textwidth}
		\centering
		\includegraphics[width=.99\linewidth]{rg_qvolve.eps} 
		
		
	\end{minipage}
	\hfill
	\begin{minipage}[b]{0.45\textwidth}
		\centering
		\includegraphics[width=.99\linewidth]{rg_xvolve.eps} 
	\end{minipage}
	\caption{ \small A comparison of gluon distributions obtained in the GLR-MQ-ZRS equation with those obtained in the GLR-MQ equation. In the figure on the left, the $R_G$ ratio is plotted with respect to $Q^2$ at $x=10^{-2}\,,10^{-4}\,\text{and}\,10^{-6}$ respectively. In the figure on the right $R_G$ ratio is plotte w.r.t $x$ at $Q^2=5,7.5,10,25,40\,\text{and}\,75\, GeV^{2}$ respectively .} 
\end{figure}

\begin{figure}[t]
	\centering
	\includegraphics[width=\textwidth]{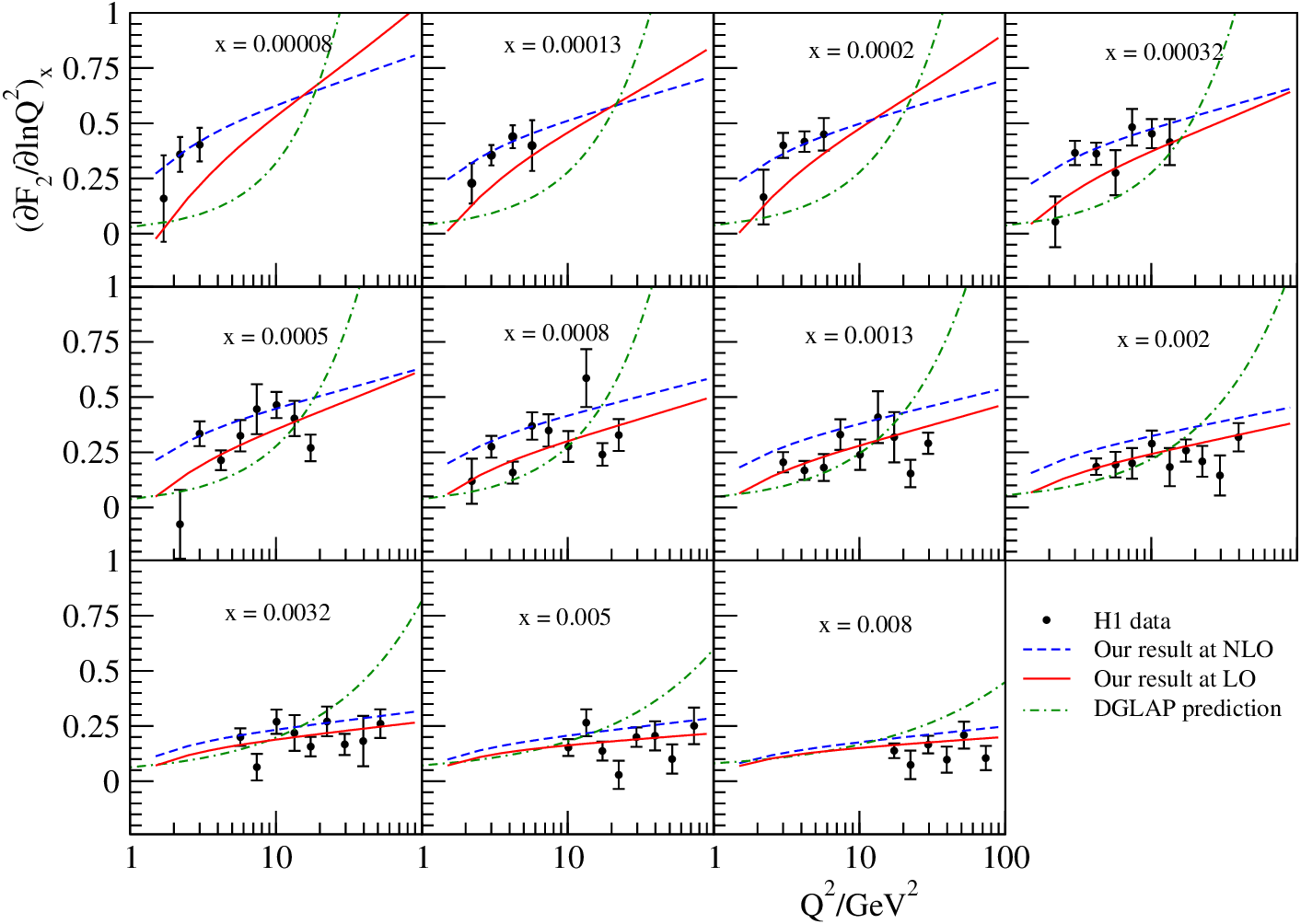}
	\caption{\small Our results of partial derivative $(\partial F_2/\partial \ln Q^2)_x$ taken at fixed x and plotted as a function of $Q^2$. The solid lines represent our prediction at LO of $P_{qg}$, whereas the dashed lines represent our prediction at NLO of $P_{qg}$ respectively. The dashed dot lines represent prediction from the solution of DGLAP equation. Our results are compared with the H1 data. The error bars represent the total errors due to both statistical and systematic errors.    }
	\label{fig:1}       
\end{figure}

\begin{figure}[t]
	\centering
	\includegraphics[width=0.8\textwidth]{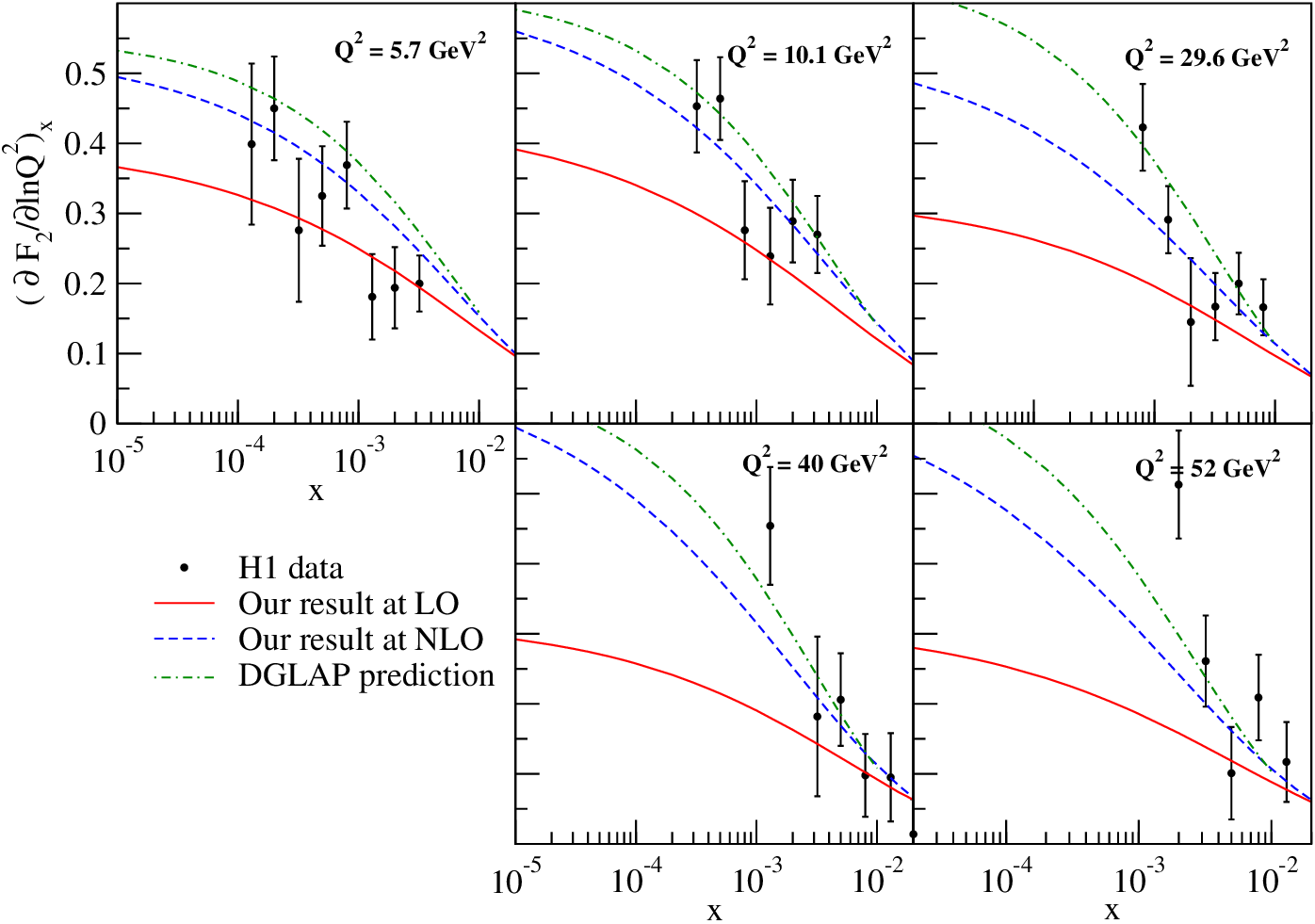}
	\caption{\small Our results of partial derivative $(\partial F_2/\partial \ln Q^2)_x$ taken at fixed $Q^2$ and plotted as a function of $x$. The solid lines represent our prediction at LO of $P_{qg}$, whereas the dashed lines represent our prediction at NLO of $P_{qg}$ respectively. The dashed dot lines represent prediction from the solution of DGLAP equation. Our results are compared with the H1 data.  The error bars represent the total errors due to both statistical and systematic errors.    }
	\label{fig:1}       
\end{figure}

\section{Results and Discussions}

In this paper we have solved the nonlinear GLR-MQ-ZRS equation and suggested  solutions in the kinematic range of $10^{-2}\leq x \leq 10^{-6}$ and $5\, GeV^2 \leq Q^2 \leq 100 \, GeV^2$ respectively. We employed Regge behaviour of gluons in the gluon rich situations. We present both the $x$ and $Q^2$ evolutions of gluon distribution function $G(x,Q^2)$. Our result of $G(x,Q^2)$ are compared with that of NNPDF3.0\cite{ball2015parton}, CT14\cite{ct14} and PDF4LHC15\cite{pdf4lhc};  and are in good agreement. The NNPDF3.0 PDF sets were determined with a methodology validated by closure test and uses global dataset including both from HERA and LHC experiments. From the HERA they have used the data of HERA-II deep-inelastic inclusive cross-sections as well as the combined charm data. They have also included jet production data from the ATLAS and CMS collaboration, vector boson rapidity and transverse momentum distributions from ATLAS, CMS and LHCb, $W+c$ data from CMS and top quark pair production total cross sections from ATLAS and CMS. On the other hand, CT14 PDF sets are the results from a global analysis by the CTEQ-TEA group. The analysis in CT14 includes measurements of inclusive vector boson  and jets production  from LHC at 7 and 8 TeV as input for the fits. Besides, they have also included the data on charm production from DIS at HERA  and precise measurements of the electron charge asymmetry from $D\varnothing$ at 9.7 $fb^{-1}$. PDF4LHC15\cite{pdf4lhc} was suggested in the year 2015 by PDF4LHC recommendations and the PDF sets are based on a statistical combination of three PDF sets viz., CT14,  MMHT2014 and NNPDF3.0 PDF sets.

\par  In order to extract $G(x,Q^2)$ from these Global PDF groups, we have used APFEL tool\cite{bertone2014apfel,carrazza2015apfel} and we have chosen LHAPDF6\cite{buckley2015lhapdf6}
  library for evolution. APFEL is a PDF evolution package written in FORTRAN 77, that allows to perform DGLAP evolution up to NNLO in QCD and to LO in QED, in the variable-flavor-number scheme (VFNS) and with either pole or $\overline{MS}$ heavy quark masses. We have also used the IMParton C++ package\cite{IMParton16,chen2014applications,zhu2016determination} to extract $G(x,Q^2)$ in the kinematic region of consideration. The IMParton package is based on the analysis to
deep inelastic scattering data applying DGLAP equations with nonlinear
corrections which gives PDFs of the proton
starting from low $ Q^2 \sim 0.07 GeV^2$. This package is basically provided with two data sets of PDFs obtained from Global analysis to DIS experimental data: the data set A is obatined from the three valence quarks non-perturbative
input; and the data set B is obtained from the non-perturbative input of three valence quarks
adding flavor-asymmetric sea quarks components. We have taken the input of $G(x,Q^2)$ from NNPDF3.0 group at  $Q_0 ^2 \approx 5\, GeV^2$ and $x_0 \approx0.01$. In our phenomenology, we have considered the QCD cutoff parameter at $\Lambda = 300\, MeV$, the correlation radius $R=5 \, GeV^{-1}$ and the Regge intercept $\lambda_G \approx 0.5$ respectively. The number of flavors in our work is considered to be four. 

For comparison with the experimental data of  $(\partial F_2 (x,Q^2)/ \partial \ln Q^2)_x$, we use the experimental data recorded with the H1 detector at HERA\cite{adloff2001ep} corresponding to an integrated luminosity of $20\, pb^{-1}$, in the precise measurement of inclusive deep-inelastic $e + p$ scattering cross section reported in the kinematic range of $1.5 \leq Q^2 \leq \, 150\, GeV^2$ and $3.10^{-5}\leq
 x \leq 0.2$ respectively.

Fig. (2) represents our best predicted results of $x$ evolution of $G(x,Q^2)$ using Regge ansatz for fixed $Q^2$. The correlation radius $R$ in this case is taken to be comparable to the size of proton i.e. $5\, GeV^{-1}$. We show the small-x behaviour of $G(x,Q^2)$ at six different values of $Q^2$ viz., 5, 10, 25, 40, 50 and 75 $GeV^2$ respectively. We observe from the figure that $G(x,Q^2)$ increases as $x$ decreases, which is in agreement with the perturbative QCD fits at small-x. We also observe that when $\alpha_{s}$ was kept fixed at 0.2(this work), $G(x,Q^2)$ rises slowly in comparison to the case when $\alpha_{s}$ had $Q^2$ dependency. Thus, $G(x,Q^2)$ for fixed $\alpha_{s}$ lies below the $G(x,Q^2)$ for $\alpha_{s} \rightarrow  \alpha_{s}(Q^2)$. Our result of $G(x,Q^2)$ are in good agreement with that of NNPDF3.0, CT14 and PDF4LHC15 in the region of $x \approx 10^{-2} - 10^{-4} $. However, the rise of gluon distributions are seemed to be tamed down towards small-x ($x \leq 10^{-4}$) in comparison to  what was predicted by NNPDF3.0, CT14 and IMParton. The x evolution of $G(x,Q^2)$ towards small-x from the IMParton package predicts strong growth of gluons, which is in contrast to our predicted result. Among the various global PDF sets, our results are more compatible with the PDF4LHC15 set. 

In Fig. (3), we present the $Q^2$ evolution of $G(x,Q^2)$ at five different values of $x$ viz., $10^{-2}$,$10^{-3}$,$10^{-4}$,$10^{-5}$ and $10^{-6}$ respectively. The gluon distribution function $G(x,Q^2)$ from our solution increases with the increase in $Q^2$, as expected. From this figure  we observe that our solution agrees very well with the predictions of NNPDF3.0, CT14, PDF4LHC and IMParton at $x= 10^{-2}$ and $10^{-3}$ respectively. However, at $x \leq 10^{-3}$, $G(x,Q^2)$ from our solution lies below the predicted $G(x,Q^2)$ from NNPDF3.0, CT14 and IMParton. From this figure it can be seen that the $Q^2$ evolution of $G(x,Q^2)$ from our solution are in better agreement with the predicted $G(x,Q^2)$ by PDF4LHC15 set. 

The sensitivities of the model parameters $\lambda_G$ and $R$ on our results are shown in terms of 3D plots in Fig. (4). At the top of Fig. (4), we plot $G(x,Q^2)$ as a function of $R$ and $Q^2$ at $x=10^{-2}$ and $10^{-6}$ respectively. At $x=10^{-2}$, no significant variation of $G(x,Q^2)$ is observed with respect to $R$ as $Q^2$ increases. But, at $x=10^{-6}$, significant increase in $G(x,Q^2)$ is observed as $R$ increases from  $2$ to $5$ $GeV^{-1}$. Therefore, we can conclude that the nonlinear effects are minimized when the gluons are spread throughout the size of the proton. This behaviour is only significant at very small values of $x$. Therefore, the model parameter $R$ becomes less significant at large values of $x$. At the bottom of Fig. (4), the $\lambda_G$ sensitivity of $G(x,Q^2)$ is shown w.r.t. $Q^2$ at $x=10^{-2}$ and $10^{-6}$ respectively. The gluon distributions are highly sensitive to $\lambda_G$ both at $x=10^{-2}$ and $10^{-6}$. As $\lambda_G$ decreases, $G(x,Q^2)$ increases steeply with the increase in $Q^2$. It can also be noted that the antishadowing effect is largest at large $Q^2$, while at smaller $Q^2$ there is very little 
effect when going from large to small x. This may be because of the fact that the transverse size of interacting gluons goes as $1/Q$. As the size of the interacting gluons are larger at small values of $Q^2$, the gluons tend to spatially overlap with each other and hence, the shadowing corrections will be more in this region. But, at large values of $Q^2$, the size of the gluons become smaller and the shadowing corrections in this region will be less and hence, the antishadowing effect will take over the shadowing effect. Therefore, the antishadowing effect is seen more towards larger values of $Q^2$. Out of the two parameters $R$ and $\lambda_G$ in our analysis, $\lambda_G$ is seen to be more sensitive to antishadowing corrections at large $Q^2$.

In Fig. (5), we study the contribution of antishadowing corrections to the overall growth of gluons by comparing the solutions of GLR-MQ-ZRS equation with the solutions of GLR-MQ equation. The parameter $R_G$ is defined as the ratio of $G(x,Q^2)$ with shadowing corrections to $G(x,Q^2)$ when both the shadowing and antishadowing corrections were considered. In Fig. 5(a),we show the $Q^2$ evolution of $R_G$ at $R=2$ and $5\, GeV^{-1}$. From the graph, we observe that for a fixed value of $x$, the $R_G$ value decreases immediately with the increase in $Q^2$, but attains a nearly constant value on further increasing $Q^2$. This means the antishadowing effect is more at low-$Q^2$, and at higher values of $Q^2$, the antishadowing effect gets balanced by the shadowing effect. This behaviour can be expected from the fact that the transverse size of gluons grows with $1/Q$. As the size of gluons increases, the probability of gluons to spatially overlap among each other increases giving rise to recombination effects. We also observe that as $x$ decreases from $10^{-2}$ to $10^{-6}$, the $R_G$ distributions shift downwards, which means the antishadowing effect is more towards smaller-x. We also observe that when gluons are spread throughout the size of the proton ($R=5\, GeV^{-1}$), the antishadowing contributions are significantly more in comparison to the situation when gluons are concentrated at hotspots($R=2\, GeV^{-1}$). In Fig. 5(b), $x$ evolution of $R_G$ distributions are plotted at six different values of $Q^2$ viz., $5$, $7.5$, $10$, $25$, $40$ and $75$ $GeV^2$ respectively. $R_G$ decreases with the decrease in $x$ for any fixed value of $Q^2$. As $Q^2$ is increased, the $R_G$ distributions shift upwards and the value of $R_G$ approaches to unity. So, at higher $Q^2$, the contributions from antishadowing corrections become significantly small.  

In Fig. (6), we plot the $Q^2$ evolution of protons $F_2$ derivative i.e. $\partial F_2 (x,Q^2)/\partial \ln Q^2$ at fixed values of $x$. Our results predict rise of $\partial F_2 (x,Q^2)/\partial \ln Q^2$ with increasing $Q^2$ which is in agreement with the experimental data recorded by H1 detector \cite{adloff2001ep} at HERA. We have included the QCD corrections of gluon-quark splitting functions upto next-to-leading order in our results. It can be observed that our LO result is in good agreement with the experimental data at large values of $x$, whereas the NLO result agrees well at the smaller values of $x$. This means towards small-x, we cannot ignore the higher order QCD contributions on the splitting kernels. We have also compared our results with the results obtained from the solution of the standard DGLAP evolution equation at LO of QCD using Regge ansatz. For both the solutions of ZRS equation and DGLAP equation, we take the same value for the parameters $\lambda_G$ and $R$. From the figure it can be seen that the $F_2$ derivative from the solution of DGLAP equation rises steeply with increasing $Q^2$. At small-x, the solution of DGLAP with Regge ansatz seems to fail miserably in describing the experimental data. Thus clearly it can be inferred that the solution of ZRS equation is more compatible with the experimental data as compared to the solution of standard DGLAP equation. 

In Fig. (7), we plot the $x$ evolution of $\partial F_2 (x,Q^2)/\partial \ln Q^2$ for fixed values of $Q^2$. Our results show a rise of $\partial F_2 (x,Q^2)/\partial \ln Q^2$ as $x$ decreases. Here, we observe that at $Q^2 = 5.7\, GeV^2$, our LO result agrees well with the experimental data; but, at subsequently higher values of $Q^2$, it is the NLO result which shows a better description of the experimental data. This suggests that for further higher values of $Q^2$, we must include  higher order QCD corrections. In this figure also it can be observed that the solution of DGLAP rises steeply as compared to the solution of ZRS equation, when $x$ decreases to small-x. The available HERA data of $\partial F_2 (x,Q^2)/\partial \ln Q^2$ with respect to $x$ in the given kinematic range is thus explained better by the solution of ZRS equation.

\section{Conclusion}

We have solved the GLR-MQ-ZRS equation using Regge ansatz and performed phenomenological study of the solutions in kinematic region of $10^{-2}\leq x \leq 10^{-6}$ and $5 \, GeV^2 \leq Q^2 \leq 100\, GeV^2$ respectively. The study was useful to validate the Regge behaviour of gluons and from our study it can be inferred that Regge behaviour of gluons provide a good description of physical picture in the kinematic range of consideration for this work. We believe that our solution is valid in the vicinity of saturation scale where nonlinear effects like gluon recombination cannot be ignored.  Using the solution of gluon distribution function we have also obtained the logarithmic derivative of $F_2$ structure function. Our result of $\partial F_2 (x,Q^2)/\partial \ln Q^2$ at leading order in $P_{qg}$ agrees well with the HERA experimental data for larger values of $x$, whereas towards smaller values of $x$, our results with the inclusion of next-to-leading order QCD corrections in $P_{qg}$ agrees significantly well. It is also interesting to conclude from our study that the antishadowing effects contribute more on gluon distribution functions towards smaller values of $x$ for fixed $Q^2$, while for fixed values of $x$ this contribution is more towards lower values of $Q^2$. At smaller-x, the antishadowing contribution increases with the increase in correlation radius $R$ among the interacting gluons; these corrections have signifcant contributions when the gluons are populated across the proton's size. Finally we conlude that among the model parameters $\lambda_G$ and $R$, our results are more sensitive to $\lambda_G$.

\par

One of the most interesting topic at small-x physics is the saturation phenomenon. Already there has been evidence of saturation from the pioneering finding of geometrical scaling in the description of experimental data at HERA\cite{geometrical_hera} , and in the production of comprehensive jets in the LHC data\cite{geometrical_lhc}; but, there has not been any conclusive proof of it yet.  It is to note that the proposed Large Hadron electron Collider (LHeC)\cite{0954-3899-39-7-075001} facility will exploit the new world of energy and intensity offered by the LHC for electron-proton scattering, through the addition of a new electron accelerator. At LHeC there is a better possibility to explore small-x physics as it is expected to reach a wider kinematic range of $x$ and $Q^2$. The phenomenological study which we have performed in this work will be very useful in the study of  saturation phenomenon at very high dense gluon regime inside the hadrons.

\par Finally, we advocate on the applicability of GLR-MQ-ZRS nonlinear equation. We believe that the GLR-MQ-ZRS equation can be a better option to study small-x phenomenon, and in extraction of PDFs. Unlike the other nonlinear equations like the GLR-MQ equation in which momentum conservation is violated, the conservation of momentum is naturally restored in the ZRS
 version of equation. Therefore, there is a greater possiblity to explore the solutions of GLR-MQ-ZRS equations and to apply them in the study of all the DIS processes. Infact for our future work, we are motivated to explore more on the GLR-MQ-ZRS equations in the study of DIS observables such as the proton's $F_2$ structure function, proton's longitudinal structure function $F_L$ and the reduced cross sections $\sigma_r$ for inclusive inelastic scattering processes $ep\rightarrow eX$, using higher order QCD corrections at higher twist level.
\section*{acknowledgments}
 One of the authors (Madhurjya Lalung) is grateful to the Council of Scientific and Industrial Research (CSIR), New Delhi, India for CSIR Senior Research Fellowship(SRF). J. K. Sarma acknowledges Council of Scientific and Industrial Research(CSIR),  New  Delhi  for  the  financial  support  in  terms  of  a  sponsored  project  with  grant  no.03(1447)/18/EMR-II.

%



\end{document}